# Data-driven Approach to Parameterize SCAN+*U* for an Accurate Description of 3*d* Transition Metal Oxide Thermochemistry

Nongnuch Artrith[1*†], José Antonio Garrido Torres[1], Alexander Urban[1], and Mark S. Hybertsen [2‡]

[1]*Department of Chemical Engineering, Columbia University, New York, NY 10027, USA.*
[2]*Center for Functional Nanomaterials, Brookhaven National Laboratory, Upton, NY 11973, USA.*
*Present Address: Materials Chemistry and Catalysis, Debye Institute for Nanomaterials Science, Utrecht University, 3584 CG Utrecht, The Netherlands.*
[†]email: n.artrith@uu.nl  [‡]email: mhyberts@bnl.gov

**Abstract:**
Semi-local density-functional theory (DFT) methods exhibit significant errors for the phase diagrams of transition-metal oxides that are caused by an incorrect description of molecular oxygen and the large self-interaction error in materials with strongly localized electronic orbitals. Empirical and semiempirical corrections based on the DFT+*U* method can reduce these errors, but the parameterization and validation of the correction terms remains an on-going challenge. We develop a systematic methodology to determine the parameters and to statistically assess the results by considering interlinked thermochemical data across a set of transition metal compounds. We consider three interconnected levels of correction terms: (1) a constant oxygen binding correction, (2) Hubbard-*U* correction, and (3) DFT/DFT+*U* compatibility correction. The parameterization is expressed as a unified optimization problem. We demonstrate this approach for 3*d* transition metal oxides, considering a target set of binary and ternary oxides. With a total of 37 measured formation enthalpies taken from the literature, the dataset is augmented by the reaction energies of 1,710 unique reactions that were derived from the formation energies by systematic enumeration. To ensure a balanced dataset across the available data, the reactions were grouped by their similarity using clustering and suitably weighted. The parameterization is validated using leave-one-out cross-validation (CV), a standard technique for the validation of statistical models. We apply the methodology to the strongly constrained and appropriately normed (SCAN) density functional. Based on the CV score, the error of binary (ternary) oxide formation energies is reduced by 40% (75%) to 0.10 (0.03) eV/atom. A simplified correction scheme that does not involve SCAN/SCAN+*U* compatibility terms still achieves an error reduction of 30% (25%). The method and tools demonstrated here can be applied to other classes of materials or to parameterize the corrections to optimize DFT+*U* performance for other target physical properties.

## 1. Introduction

Density-functional theory (DFT) [1,2] has become a standard tool for computational materials design.[3–6] However, conventional (semi-)local DFT methods based on the generalized gradient approximation (GGA)[7] exhibit a self-interaction error (SIE) that can lead to an over-delocalization of electrons, resulting in an incorrect description of many TM oxides with strongly localized $d$ electrons.[8–10] In addition, the widely used GGA functional by Perdew, Burke, and Ernzerhof (PBE)[11] overestimates the strength of the O–O bond in the dioxygen molecule, which introduces an additional error in the formation energies of oxides.[12] For applications involving TM oxides, empirical corrections for both the SIE and the overbinding of oxygen are therefore often introduced,[13–17] the values of which have to be carefully fitted and validated.

Accounting for strong, local Coulomb interactions between electrons in the TM $d$ states approximately through the DFT+$U$ method has been particularly successful for reducing the SIE in TM oxides without significantly increasing the computational cost.[13,14] DFT+$U$ introduces TM-specific Hubbard $U$ parameters representing the local, screened Coulomb interaction. They can be either obtained from linear-response theory[18,19] or empirically by fitting reference properties such as reaction energies or band gaps from experiment or more accurate electronic structure calculations.[14] In combination with an empirical energy correction for the overbinding of the O-O bond[15] and a compatibility correction that accounts for the mixing of PBE and PBE+$U$ calculations,[20] PBE/PBE+$U$ often reproduces TM oxide formation energies with sufficient accuracy so that derived phase diagrams are in agreement with experiment.[21]

One challenge in the parameterization of DFT+$U$ methods is the dependence of the optimal Hubbard $U$ parameter on the oxidation state of the metal center.[22,23] Energies obtained from DFT+$U$ calculations with different $U$ values lack a common reference and are therefore not generally compatible. In practice, energies associated with oxidation or reduction reactions must be calculated with average $U$ values that are a compromise for all involved valence states.[23] DFT+$U$ is a static correction to DFT and does not capture, for example, frequency-dependent screening by delocalized electrons,[19] which would call for more complex theories, such as DFT+DMFT.[24,25] Furthermore, the convergence of DFT+$U$ calculations to the self-consistent electronic ground state becomes more challenging with increasing $U$ values.[21,26]

While some of the challenges of DFT+$U$ are intrinsic to the approach, the choice of exchange-correlation functional also plays an important role for the accuracy that DFT+$U$ calculations can achieve. For example, PBE+$U$ predicts an incorrect hybridization of TM and O states for oxides in which the TM $d$ states are close in energy to the O $2p$ states.[22] This electronic-structure error cannot be removed with an empirical energy correction and requires an additional electronic-structure correction, such as an additional Hubbard-$U$ term for the O $2p$ states.[27–29] Intuitively, a functional that predicts oxide formation energies more accurately than PBE should also provide a more robust starting point for the modeling of transition-metal oxides with the addition of empirical corrections.

The recently proposed strongly constrained and appropriately normed (SCAN) meta-GGA functional has shown promise for the prediction of oxide phase diagrams with greater accuracy and less empiricism than GGA functionals.[30–39] Prior studies concluded that SCAN does not entirely remove the SIE, and a Hubbard-$U$ correction is still required for many 3$d$ TM species, albeit with the magnitude of the $U$ values being smaller than required for PBE in applications considered to date.[32–36,38–42] The performance of SCAN/SCAN+$U$ for the prediction of TM oxide *formation energies* has, to our knowledge, not yet been investigated.

In the present work, we determine the parameters for the three levels of corrections needed to facilitate quantitative formation-energy calculations with SCAN/SCAN+$U$: (i) the O–O binding energy correction, (ii) the Hubbard-$U$ electronic-structure correction of the TM $d$ states, and (iii) the SCAN/SCAN+$U$ compatibility correction. To accomplish this parameterization in a systematic and unbiased fashion, we propose a methodology for the automated fit of Hubbard-$U$ values and DFT/DFT+$U$ compatibility corrections to experimental formation energies from the literature and derived reaction energies. By considering a large, interconnected set of compounds simultaneously, we are able to apply quantitative metrics that give a statistical outlook for the derived parameters.

Specifically, we demonstrate and apply the framework to determine an optimal parameterization of SCAN+$U$ for the prediction of the formation energies of binary and ternary oxides of the 3$d$ TM species Ti, V, Cr, Mn, Fe, Co, Ni, and Cu, as well as energies of reactions involving the oxide species. We report benchmarks of reaction and formation energies that were not included in the optimization procedure and assess the impact of the +$U$ correction on the prediction of oxide phase diagrams. The reference data set contains 37 formation energies and 1,710 oxide reaction energies.

Our methodology optimizes $U$ values for a family of transition metal oxide reactions and formation reactions incorporating robust statistical methods that remove arbitrariness from the determination of empirical corrections in DFT+$U$ approaches. It provides error estimates both for the parameters and for the prediction of reaction energies, including the impact of transferability of the parameters. Key elements of the approach include regularized least-squares optimization, cross-validation, and grouping of similar oxide and formation reactions using principal component analysis and $k$-means clustering to remove biases from the reference data set. Leveraging this framework and the new SCAN functional yields a SCAN/SCAN+$U$ parameterization for accurate calculations of 3$d$ TM metal and oxide reactions.

In Section 2, we describe the theoretical background and the essential features of our new framework. The main results are presented in Section 3 and further discussion appears in Section 4. We summarize and conclude the paper in Section 5.

## 2. Methods

### 2.1. Terminology and definitions

The formation of a binary TM oxide $A_xO_y$ (A = TM species) from the elemental metal and oxygen gas is described by the general *formation reaction*

$$x\text{A} + \frac{y}{2}\text{O}_2 \rightarrow A_xO_y \ . \tag{1}$$

The enthalpy of formation $\Delta_f H(A_xO_y)$ is defined as the heat of reaction of the formation reaction, *i.e.*, the enthalpy difference between the products and reactants in equation (1). At zero Kelvin, ignoring corrections due to zero-point fluctuations, the enthalpy of formation can be approximated as the difference of the DFT energies ($E_{\text{DFT}}$) of the reactants and products

$$\Delta_f H(A_xO_y) \approx \Delta_f H_{\text{DFT}}(A_xO_y) = E_{\text{DFT}}(A_xO_y) - x\, E_{\text{DFT}}(A) - \frac{y}{2} E_{\text{DFT}}(O_2) \ . \tag{2}$$

The definition for ternary oxides is equivalent, and the enthalpy of formation of a ternary oxide $A_xB_yO_z$ is given by

$$\Delta_f H_{\text{DFT}}(A_xB_yO_z) = E_{\text{DFT}}(A_xB_yO_z) - x\, E_{\text{DFT}}(A) - y\, E_{\text{DFT}}(B) - \frac{z}{2} E_{\text{DFT}}(O_2) \ . \tag{3}$$

In the following, we refer to zero-Kelvin formation enthalpies simply as *formation energies*.

The heat of reaction $\Delta_r H$ of a general reaction can be expressed in terms of formation energy differences of reaction products $P$ and reactants $R$

$$\Delta E = \Delta_r H = \overset{\text{Products}}{\underset{P}{\sum}} \Delta_f H(P) - \overset{\text{Reactants}}{\underset{R}{\sum}} \Delta_f H(R) \ , \tag{4}$$

which we refer to as *reaction energy* $\Delta E$ in the following.

In addition to formation reactions, we also consider the reaction energies of solid-state reactions. Here, we distinguish between pure *oxide reactions not involving elemental oxygen* ($O_2$) and *oxide reactions involving $O_2$*. Oxide reactions not involving $O_2$ are reactions in which all reactants and products are oxides, such as

$$A_xO_{z_1} + B_yO_{z_2} \rightarrow A_xB_yO_{z_1+z_2} \ . \tag{5}$$

Such oxide reactions do not typically involve the change of oxidation states, and we expect therefore that a single $U$ value can be found that is simultaneously (close to) optimal for the reactants and products. In contrast, in oxide reactions that involve elemental oxygen as either reactant or product at least one TM species has to be subject to a change of oxidation state, as elemental oxygen has an oxidation state of zero whereas the formal oxidation state of oxygen in oxides is –2

$$A_xO_y + \frac{1}{2}O_2 \rightarrow A_xO_{y+1} \ \ \text{with} \ y > 0. \tag{6}$$

Calculating the reaction energies of such reactions might require an average $U$ value that is neither optimal for the reactants nor for the products, since the optimal Hubbard $U$ parameter can vary with the oxidation state. [18,22]

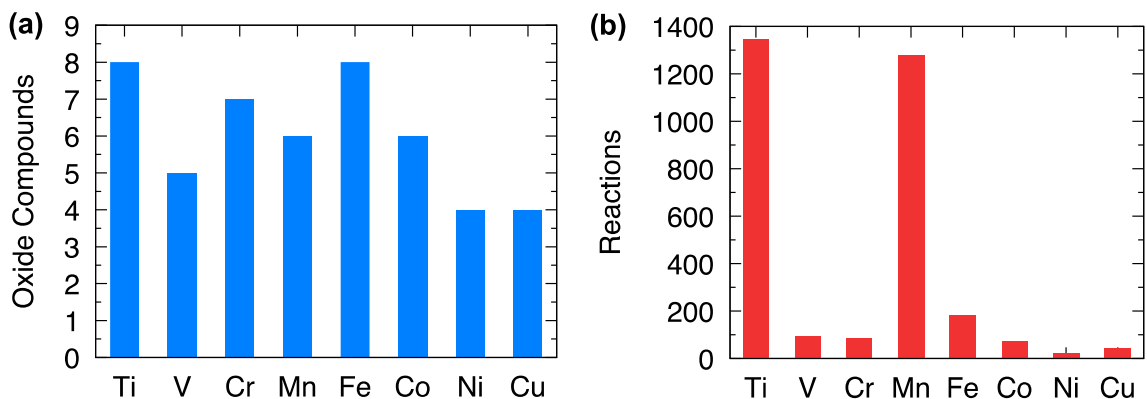

**Figure 1** Representation of different transition-metal species in our data set. **(a)** Number of oxide compounds with formation enthalpies in our reference data set. **(b)** Number of derived oxide and $O_2$ reactions. See **Tables S1 and S2** for a complete list.

## 2.2. Reference data

Our database contains the experimental formation enthalpies of 24 binary oxides ($A_xO_y$) and 13 ternary oxides ($A_xB_yO_z$) of $3d$ TMs from the literature.[20,43–45] Our calculations correspond to 0 K and 0 atm, but we chose reference enthalpies at standard conditions (298 K and 1 atm) for which more data is available. The enthalpy difference $\Delta_f H^{0K} - \Delta_f H^{298K}$ has previously been estimated to be typically less than 0.03 eV/atom,[20,46] so that this error is not significant. See **Table S1** for a complete list of compounds and the original references. We did not consider a Hubbard-$U$ correction for $Sc_2O_3$ and ZnO, since $Sc^{3+}$ and $Zn^{2+}$ have an empty and filled $3d$ band, respectively. In such cases, Hubbard-$U$ corrections are usually not needed. As an important validation of this assumption, our computations with the uncorrected SCAN functional accurately predict the formation enthalpies of ZnO and $Sc_2O_3$ with errors of 0.045 eV/$O_2$ and 0.026 eV/$O_2$, respectively (see also supplementary **Table S3**). The frequency of occurrence of the other $3d$ TMs in our reference data set is visualized in **Figure 1a**.

From the data set of $24 + 13 = 37$ experimental formation enthalpies, the energies of oxide reactions (with and without $O_2$ as reactant) can be derived according to equation (4). We used the python materials genomics (pymatgen)[47] package for enumerating all unique oxide reactions involving the 37 oxide compounds, yielding a derived database of 1,710 unique reaction energies (**Table S2**). Note that the frequency of occurrence of the different TM species in the derived reaction database varies, as seen in **Figure 1b**. Ni is represented by the smallest number of reactions (seven). On the other hand, Ti, Mn, and Fe form oxides with three different oxidation states and measured formation energies are available for more compounds. The derived reaction data set therefore contains a large number of reactions involving these three species. The complete list of reactions and their derived energies are given in supplementary **Table S2**.

In the analysis of the O–O binding energy correction, we also consider a sample of other oxides ($Al_2O_3$, $CaO$, $Li_2O$, $MgO$, $Na_2O$, $Sc_2O_3$, $SiO_2$, $SnO_2$, $ZnO$, $ZrO_2$), see also **Table S3**.

### 2.3. Grouping of similar reactions

The enumeration procedure described in the previous section produces a complete set of reactions, but for some species, many reactions have similar reactants. This could potentially lead to an overrepresentation in the reference data set. Note that our optimization methodology employs regularization (described below) to make the results more robust with respect to biases in the reference data set. However, in addition, we propose a metric for reaction similarity that was used to group similar reactions and thus determine a shared weight for the reactions in the optimization.

To determine reaction similarities systematically, each chemical reaction was represented as a vector with 57 components, each of which represents the coefficient of one of the compounds in our library (37 oxides and their base elements). We employ the convention that reactant coefficients are negative and product coefficients are positive, so that the reaction energy of equation (4) can be expressed as the scalar product $\Delta E = \vec{r}\,\vec{e}_U$, where $\vec{r}$ is the vector with reaction coefficients and $\vec{e}_U$ is a vector with the computed compound energies for a specific set of $U$ values. In this representation, a grouping of similar reactions was achieved in two steps by (1) reducing the dimension of the reaction representation via principal component analysis (PCA)[48] and (2) performing a cluster analysis using *k*-means clustering.[49] A reference to the Python code implementing the reaction grouping methodology (using *scikit-learn*[50]) is given in the code availability section below. Reactions that are assigned to the same cluster are considered similar in our optimization procedure and enter with a shared weight. This means, all $N_i$ reactions within a cluster $i$ enter the optimization procedure with a weight of $1/N_i$.

### 2.4. Empirical corrections of DFT errors

As described in the introduction, we considered corrections to three sources of errors in DFT formation energy calculations in the present work: (i) O–O overbinding, (ii) the SIE of DFT, and (iii) the DFT/DFT+$U$ incompatibility. In this section, we express the parameterization of the three empirical correction terms as formal optimization problems.

O–O overbinding affects the prediction of reaction energies that involve elemental oxygen, *i.e.*, formation energies and $O_2$ reaction energies. We employ the technique by Wang et al.[15] for determining a constant correction to the energy of the $O_2$ molecule. In this approach, the systematic error in oxide formation energies is determined by comparison of predicted formation energies from DFT calculations and the corresponding reference energies from tabulated experiments.

To express the oxygen correction by Wang et al. as an optimization problem, we introduce the *objective function* (or *loss function*)

$$L_{O_2} = \sum_i \left( \left[ E_{\text{DFT}}(\sigma_i) + \frac{n_O(\sigma_i)}{2} \varepsilon_{O_2} \right] - \Delta_f H(\sigma_i) \right)^2, \qquad (7)$$

where $\sigma_i$ is an oxide composition and $n_O(\sigma_i)$ is the number of oxygen atoms in $\sigma_i$. The optimal $O_2$ correction energy $\varepsilon_{O_2}^{corr}$ is then determined by minimizing the objective function

$$\varepsilon_{O_2}^{corr} = \arg\min_{\varepsilon_{O_2}} L_{O_2} \ . \tag{8}$$

For the correction of the SIE, we employ a rotationally invariant Hubbard-$U$ term[14] for the TM $d$ bands. Since our objective is the accurate prediction of formation energies and phase diagrams, we follow here also the approach by Wang et al.[15] and adjust the $U$ values such that experimental reaction energies are reproduced as well as possible. The objective function for the $U$-value optimization is therefore

$$L_U = \sum_i \left(\Delta E_{DFT+U}(\rho_i, \{U\}_{TM}) - \Delta E(\rho_i)\right)^2 \ , \tag{9}$$

where $\rho_i$ is a reaction and $\Delta E_{DFT+U}(\rho_i, \{U\}_{TM})$ is the reaction energy of $\rho_i$ predicted by DFT+$U$ calculations using the set of $U$ values $\{U\}_{TM}$. The optimal $U$ values are those that minimize the objective function $L_U$.

We employ the Hubbard-$U$ correction only for the TM $d$ bands in TM oxides and not for the elemental metals, since electrons in metals are delocalized and are already well described by uncorrected local and semi-local DFT.[51] This means, the reaction energies of oxide reactions can be calculated consistently with DFT+$U$, since TM species only occur in oxide form. However, formation reactions involve both elemental (metallic) TM species, which are best described by uncorrected DFT, and their oxides, which require DFT+$U$.[20]

To ensure that the energies from DFT and DFT+$U$ calculations are compatible, we explore two different strategies: **(1)** We determine the optimal set of $U$ values by minimizing $L_U$ for a reference data set containing oxide reactions, both those involving and those not involving $O_2$, and formation reactions. In this approach, DFT/DFT+$U$ compatibility is implicit to the objective function, and the optimal $U$ values will allow calculation of energy differences by combining DFT and DFT+$U$ calculations as needed for formation energies. **(2)** We employ an additional DFT/DFT+$U$ compatibility correction following Jain et al.[20] for the calculation of formation energies.

The DFT/DFT+$U$ correction by Jain et al.[20] introduces another set of TM-species dependent correction parameters $\{\mu\}_{\text{TM}}$ that are applied to all energies from DFT+$U$ calculations, yielding *renormalized* DFT+$U$ energies

$$E_{\text{DFT}+U}^{\text{renorm}}(\sigma) = E_{\text{DFT}+U}(\sigma) - \sum_{M} n_M(\sigma)\,\mu_M \tag{10}$$

where the sum runs over all considered TM species $M$ and $n_M(\sigma)$ is the number of atoms of species $M$ in composition $\sigma$. With this compatibility correction, formation energies can be calculated by combining DFT calculations of the metals and renormalized DFT+$U$ calculations of the oxides, e.g., for binary oxides

$$\Delta_f H_{\text{DFT/DFT}+U}(M_xO_y) = E_{\text{DFT}+U}^{\text{renorm}}(M_xO_y) - xE_{\text{DFT}}(M) - \frac{y}{2}E_{\text{DFT}}(O_2) \tag{11}$$

$$= E_{\text{DFT}+U}(M_xO_y) - x\,\mu_M - x\,E_{\text{DFT}}(M) - \frac{y}{2}E_{\text{DFT}}(O_2)\;.$$

An objective function for the determination of the parameters $\{\mu\}_{\text{TM}}$ can be formulated as

$$L_\mu = \sum_i \left(\Delta_f H_{\text{DFT/DFT}+U}(\sigma_i, \{U\}_{\text{TM}}, \{\mu\}_{\text{TM}}) - \Delta_f H(\sigma_i)\right)^2 \tag{12}$$

where the sum runs over oxide compositions $\sigma_i$. Note that the value of the DFT/DFT+$U$ compatibility correction parameters $\{\mu\}_{\text{TM}}$ depends on the choice of the $U$ values. Hence, the optimization of $U$ parameters with the objective function $L_U$ of equation (9) and the calculation of the compatibility correction parameters has to be done simultaneously.

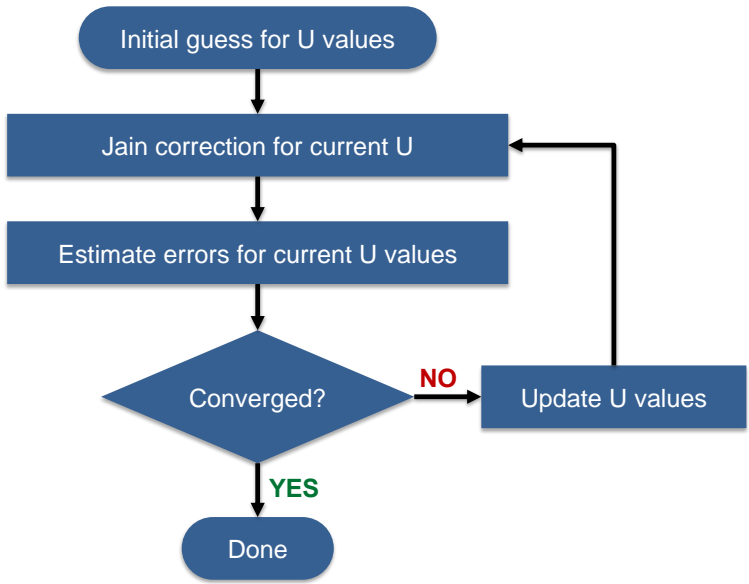

**Figure 2** Flowchart of the iterative approach for the simultaneous optimization of Hubbard-$U$ values and DFT/DFT+$U$ compatibility (Jain) corrections.

In the scope of the reactions in our dataset, the Jain correction specifically affects the formation energies where the elemental TM crystals are a reactant. It does not affect the oxide reactions, independent of whether $O_2$ is a reactant.

### 2.5. Optimization procedure

We adopted the following systematic procedure for the parameterization of the empirical correction terms described in the previous section:

First, the $O_2$ correction energy $\varepsilon_{O_2}^{corr}$ for SCAN was obtained as described in equation (8), *i.e.*, by minimizing the objective function $L_{O_2}$. The $O_2$ correction is purely a correction of the energy of the $O_2$ molecule as predicted by DFT, and it is therefore independent of the choice of $U$ values.

Next, $U$ values were determined via iterative least-squares optimization, as schematically shown in the flowchart of **Figure 2**. The optimization of $U$ values was performed by minimizing the objective function $L_U$ of equation (9) with the two strategies described in the previous sections, i.e., (1) by fitting the entire database including oxide reaction energies and formation energies, and (2) by fitting only oxide reaction energies and introducing an additional SCAN/SCAN+$U$ compatibility correction following Jain et al. (*Jain correction*).[20]

### 2.5.1 Interpolation of DFT+$U$ energies

To facilitate the rapid evaluation of the objective function for different $U$ values, the energy of a compound $E_{\text{DFT}+U}(\sigma_i)$ for a given set of $U$ values was approximated by linear interpolation based on DFT calculations for at least three different $U$ values per TM species and compound.

It has previously been established that DFT+$U$ reaction energies are, in good approximation, proportional to the $U$ value over the relevant range.[15,34] The DFT+$U$ energies of binary oxides were estimated by linear interpolation by Wang et al.[15] for PBE+$U$ and by Gautam et al.[34] and Long et al.[39] for SCAN+$U$ parameterization. Here, we generalize this approach also to ternary oxides, i.e., we determine the planes that approximate best the SCAN+$U$ energies of ternary oxides as function of the $U$ values of two TM species.

### 2.5.2 Regularization of the $U$ values

The main objective of the $U$-value optimization procedure is to determine those $U$ values that best reproduce the experimental reaction energies. To remove unnecessary flexibility from the regression, we introduce another condition: If two sets of $U$ values yield overall equivalent accuracy, $U$ values with smaller magnitude should be preferred.

To bias the optimization to smaller $U$ values, we make use of L2 regularization[52] and introduce an additional term into the objective function $L_U$ of equation (9) that scales with the norm of the $U$ values

$$L_U^{\text{reg}} = L_U + \lambda \sum_M |U_M|^2 \quad . \tag{13}$$

Determination of λ is discussed in section 3.2.

### 2.5.3 Validation and error quantification

While the goodness-of-fit metric provides a criterion for the overall accuracy of each version of the DFT+$U$ scheme across the data set, leave-one-out cross-validation (LOOCV) is a standard error quantification technique for statistical models that provides a much more robust validation criterion.[53,54] Here, we estimate the accuracy of our SCAN+$U$ parameterizations LOOCV. In practice, this means that the optimization procedure of **Figure 2** is repeated 37 times. Each time, one of the compounds $\sigma_i$ with $1 \leq i \leq 37$ selected from the set of 24 binary and 13 ternary oxides is removed from the reference data set along with all reactions in which it occurs. For the remaining 36 compounds and the associated derived reactions, the full analysis procedure is carried out, including grouping the reactions via PCA and $k$-means clustering followed by the parameterization optimization. The resulting SCAN+$U$ parameters are then used to evaluate the prediction error for the formation energy of compound $\sigma_i$ and the reaction energies of all reactions involving compound $\sigma_i$. Hence, all predicted SCAN+$U$ formation and reaction energies reported in the following are true predictions of reaction energies outside the reference data set entering the least-squares optimization.

Another benefit of the LOOCV method is that it provides a sensitivity analysis for the obtained Hubbard-$U$ and Jain compatibility correction values. Leaving out compound $\sigma_i$ from the optimization affects the optimal $U$ values for the TM species in $\sigma_i$. Thus, from LOOCV we obtain a distribution of $U$ values for each TM species. The spread of values indicates how sensitive the $U$ value is with respect to the chosen reference data set.

## 2.6. Computational details: Density-functional theory calculations

All DFT calculations were performed within the projector-augmented wave (PAW) formalism[55] as implemented in the Vienna Ab Initio Simulation Package (VASP).[56–59] VASP input files were generated using the Python Materials Genomics (pymatgen) toolkit.[47] All calculations employed the strongly constrained and appropriately normed (SCAN) meta-GGA exchange-correlation functional[30] and the rVV10 dispersion correction.[31,60] A plane wave energy cutoff of 520 eV was employed for the representation of the wave functions, and automatically generated regular Γ-centered $k$-point meshes with a length parameter of $R_k = 25$ were employed for the integration of the Brillouin zone. The convergence threshold for self-consistent field calculations was $10^{-5}$ eV, and geometry optimizations minimized the atomic forces to less than 0.1 eV Å$^{-1}$. All DFT calculations were spin polarized, and both ferromagnetic and antiferromagnetic configurations were considered in systems with unpaired electrons. For binary oxides, antiferromagnetic spin orderings were systematically enumerated using the methods implemented in enumlib[61–64] as made available in pymatgen. For ternary oxides, only a manually chosen spin orderings were considered owing to the large number of configurational degrees of freedom. For the O–O binding energy analysis of **Figure 3**, DFT calculations with the GGA functional by Perdew, Burke, and Ernzerhof (PBE)[11] were performed using parameters identical to those of the SCAN+rVV10 calculations.

## 3. Results

### 3.1. Correction for systematic oxygen energy error

**Figure 3** shows the formation energies of binary and ternary oxides as predicted by DFT calculations compared to the experimental reference values. The details of our DFT calculations are given in section 2.6. As previously reported,[15] the PBE functional systematically underestimates the magnitude of the formation energies owing to the O–O overbinding error in $O_2$ as demonstrated by the offset in the correlation plot (**Figure 3a**). In contrast, the correlation plot for the SCAN functional (**Figure 3b**) does not exhibit any offset indicating no systematic error. Thus, no oxygen energy correction is required for the SCAN functional, and $\varepsilon_{O_2}^{corr} = 0$ eV, as also found previously in references.[34,38,39]

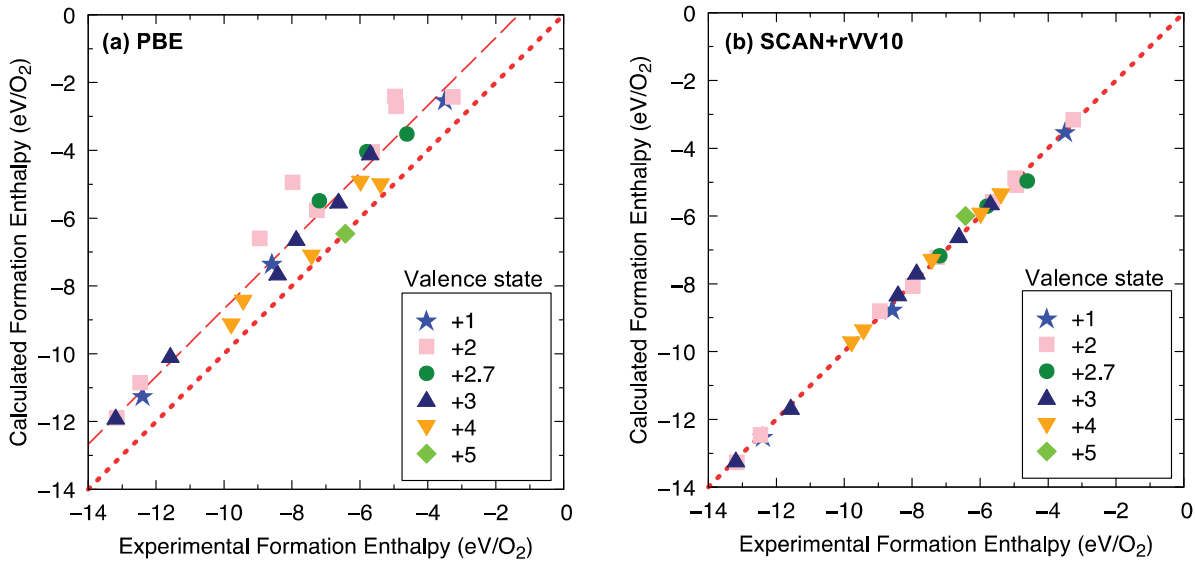

**Figure 3.** Comparison of **(a)** PBE and **(b)** SCAN+rVV10 formation enthalpies with experimental reference values. In contrast to PBE, SCAN+rVV10 formation enthalpies do not show any systematic shift due to oxygen overbinding. No empirical oxygen energy correction is needed for SCAN+rVV10. As noted in section 2.2, a sample of additional main-group oxides are included. See supplementary **Table S3**.

### 3.2. Impact of regularization

To determine a suitable value for the regularization parameter λ of equation (13), we performed a Hubbard-$U$ value optimization as described above with values of λ varying from 0.00 to 0.05.

**Figure S1** shows how the RMSE of the predicted reaction energies, the $U$ values, and the DFT/DFT+$U$ compatibility correction change with the regularization parameter. As $\lambda$ increases from 0.00 to 0.01, the increase of the RMSE is negligible, but the the $U$ values of some of the TM species are already significantly lowered. Specifically, the change of the $U$ values of Ti, Cr, and Co decrease by ~0.5 eV. Increasing the L2 regularization parameter further results in a notable increase of the RMSE for values of $\lambda$ beyond 0.01 eV (see **Figure S1**). Therefore, we chose $\lambda = 0.01$ for the analysis in the remainder of this work.

### 3.3. Reaction similarity

We performed a PCA of the 1,710 reactions in our reference data set in an initial vector representation with 57 dimensions, as described above. As shown in **Figure S2**, 25 principal components (PCs) can explain >95% of the variance of the data set. Reactions that share reactants and/or products have similar PC representations, whereas the reaction type (e.g., formation reactions vs. oxide reactions) does not strongly correlate with the PCs, as seen for the example of the first two PCs in **Figure S3**.

To group reactions by similarity in the PC representation, we next performed a *k*-means cluster analysis. The *k*-means clustering method requires that the number of clusters (i.e., *reaction groups* here) be defined beforehand. **Figure S4** shows how the optimal $U$ value, the value of the DFT/DFT+$U$ (Jain) correction, and the RMSE of the reaction energies vary with the number of clusters, which were chosen as multiples of the number of oxides in our reference data set (37, 74, 111, 148, 222, 444, 666, 888, 1110, 1332, 1554, and 1700). The impact of grouping the reactions on the optimal U value emerges as the number of clusters is reduced from 1700. As seen in the figure, the optimal $U$ value is minimally affected by the number of clusters down to about 444, with the exception of Cr where it decreases from ~4 eV to ~3 eV. Not surprisingly, when smaller numbers of clusters are employed, 222 and below, the optimal *U* values for several elements show more significant variations. Overall, at intermediate numbers of clusters, there is a broad plateau in the optimal *U* values. Based on this analysis, we chose 888 clusters (= 24×37), and all reactions were weighted according to the size of their reaction group in the *U*-value optimizations reported in the following.

### 3.4. Hubbard-*U* values for different fitting strategies

We performed *U*-value optimizations using the two strategies detailed under the theoretical framework in the previous section, *i.e.*, with and without DFT/DFT+*U* compatibility (Jain) corrections. In both cases, the iterative least-squares fit was first performed without regularization and then with L2 regularization (with $\lambda = 0.01$) as described above. No oxygen energy correction was introduced.

**Table 1.** *U* values and Jain DFT/DFT+*U* compatibility corrections obtained for different optimization strategies with and without L2 regularization. All *U* values and SCAN/SCAN+*U* corrections are given in electronvolts (eV). The recommended values are shown in bold.

|  | Ti | V | Cr | Mn | Fe | Co | Ni | Cu |
|---|---|---|---|---|---|---|---|---|
| **Optimized *U* values without SCAN/SCAN+*U* correction** | | | | | | | | |
| *U* value | 0.60 | 0.95 | 1.09 | 1.49 | 0.91 | 1.51 | 0.59 | 0.00 |
| **L2-regularized optimized *U* values without SCAN/SCAN+*U* correction** | | | | | | | | |
| *U* value | 0.54 | 0.93 | 1.02 | 1.42 | 0.86 | 1.24 | 0.38 | 0.00 |
| **Optimized *U* values with SCAN/SCAN+*U* correction after Jain *et al.*** | | | | | | | | |
| *U* value | 2.54 | 0.95 | 5.01 | 2.23 | 2.01 | 2.84 | 0.42 | 0.00 |
| Jain correction | 1.30 | 0.05 | 2.23 | 0.68 | 0.82 | 0.74 | -0.15 | 0.02 |
| **L2-regularized optimized *U* values with SCAN/SCAN+*U* correction after Jain *et al.*** | | | | | | | | |
| *U* value | **1.87** | **0.93** | **2.86** | **1.99** | **1.88** | **2.08** | **0.41** | **0.00** |
| Jain correction | **0.89** | **0.04** | **1.01** | **0.58** | **0.74** | **0.50** | **-0.15** | **0.02** |

The obtained *U* values are given in **Table 1**. Significant differences of more than 1 eV are seen between *U* values optimized with and without Jain corrections. Without Jain corrections, the *U* values for Ti, Cr, Mn, Fe, and Co are smaller, while that for Ni increases. By design, L2 regularization leads to smaller *U* values. Without Jain correction, the differences due to the L2 regularization are generally small and range from <0.1 eV to ~0.3 eV. When Jain correction energies are used, L2 regularization has a more notable impact, and the *U* values for Cr and Co decrease from 5.01 to 2.86 eV and from 2.84 to 2.08 eV, respectively. In general, the *U* values optimized for SCAN are already of modest scale.

### 3.5. DFT/DFT+*U* compatibility (Jain) corrections

**Table 1** also lists the DFT/DFT+*U* compatibility (Jain) corrections of the different TM species corresponding to the *U* values determined with strategy (2). As seen in the table, the Jain correction energies obtained from L2-regularized optimization are generally smaller than those obtained without regularization. The Jain corrections are largest in absolute value for those TM species for which the *U* values vary most among the different optimization strategies (Ti, Cr, Mn, Fe, and Co). The sign of the corrections is positive with the exception of a small negative correction for Ni (-0.15 eV).

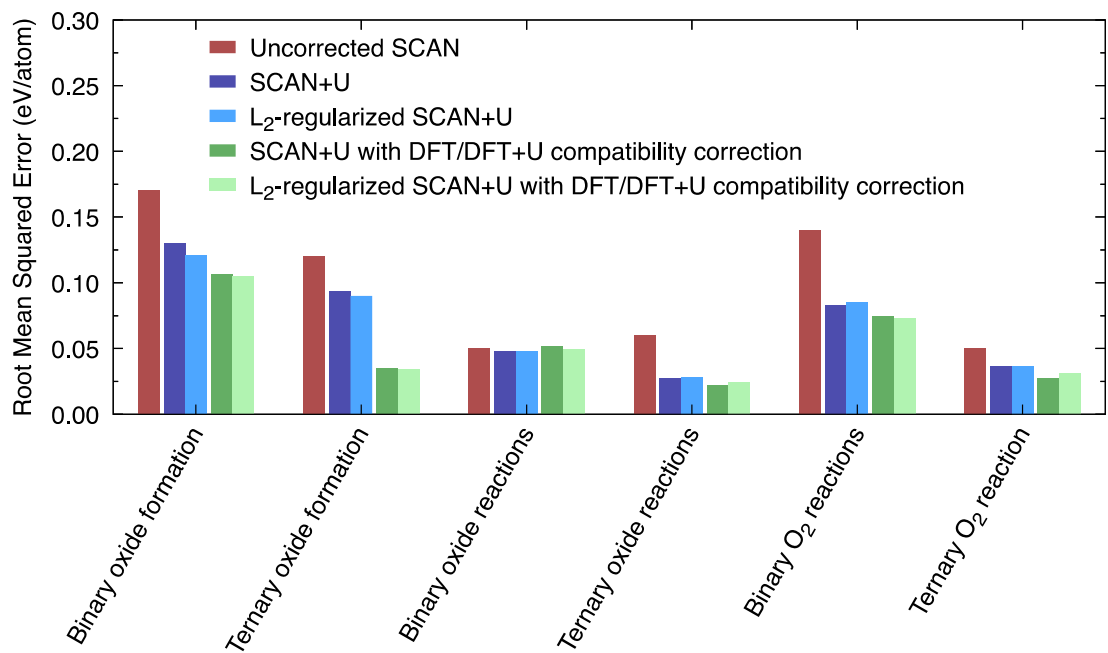

**Figure 4** Root mean squared errors for oxide formation and oxide reaction energies predicted by SCAN+*U* calculations using the three different sets of *U* values and Jain corrections of **Table 1**. Oxide reactions not involving elemental oxygen as a reactant are labeled *oxide reactions*, whereas those reactions that do involve oxygen are labeled *$O_2$ reactions*. All SCAN+*U* error estimates were obtained from leave-one-out cross-validation.

### 3.6. Validation and accuracy estimate of the different parametrizations

**Figure 4** shows the root mean squared errors (RMSE) of oxide formation energies and oxide reaction energies relative to the experimental reference data as obtained from LOOCV. See also supplementary **Table S4** for a list of the cross-validation errors.

As expected, the overall RMSE decreases as Hubbard-*U* and Jain corrections are introduced. However, the error does not decrease equally for all classes of reactions. The accuracy of the binary oxide reactions not involving $O_2$ does not improve when a Hubbard-*U* correction is introduced, and only a small improvement is seen for ternary oxide reactions that involve $O_2$. Significant improvement is seen for both binary and ternary oxide formation energies and for the energies of binary oxide reactions involving $O_2$. The error decreases from 0.17 eV/atom to 0.10 eV/atom and from 0.12 eV/atom to 0.03 eV/atom for binary and ternary formation energies, respectively. The error of the reaction energies of binary oxides involving $O_2$ improves from 0.14 eV/atom to 0.07 eV/atom. Interestingly, the Hubbard-$U$ correction as well as the additional DFT/DFT+$U$ compatibility correction after Jain both contribute to reducing the error in these cases. However, a significant impact can be realized without the Jain correction.

The good predictive power of SCAN+$U$ with the $U$+Jain+L2 parametrization for formation energies can also be seen in **Figure 5**, in which the computed formation energies are compared with their experimental references. As seen in the figure, the LOOCV predictions are centered around the optimal diagonal for binary oxides (**Figure 5a**) and ternary oxides (**Figure 5b**), respectively. The most noticeable remaining error, with a magnitude of ~0.2 eV, is seen for the binary oxide TiO.

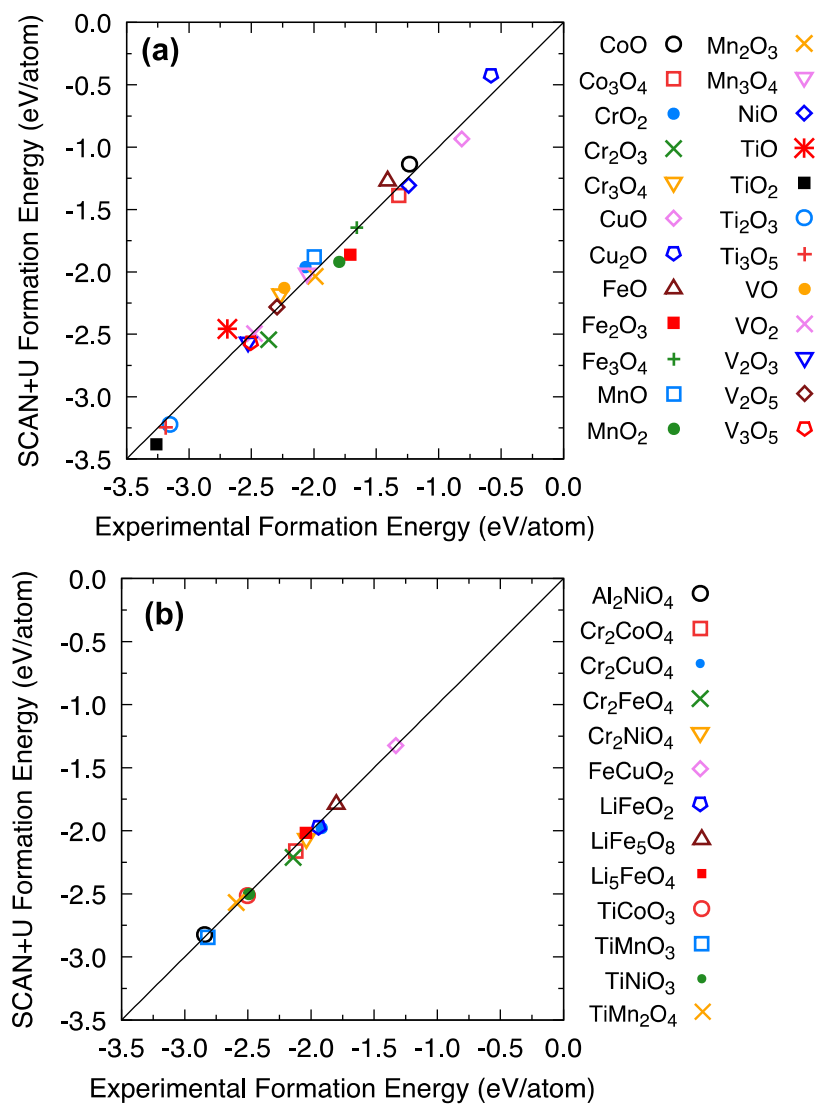

**Figure 5** Comparison of predicted enthalpies of formation ($\Delta_f H°$) from SCAN+$U$ with experimental reference values (298K) for **(a)** binary and **(b)** ternary transition-metal oxides. SCAN+$U$ predictions are based on the L2-regularized optimized $U$ values and DFT/DFT+$U$ compatibility corrections of **Table 1**. All data points shown were evaluated with leave-one-out cross-validation. Each point represents a true prediction based on the $U$-values fit to the rest of the data set.

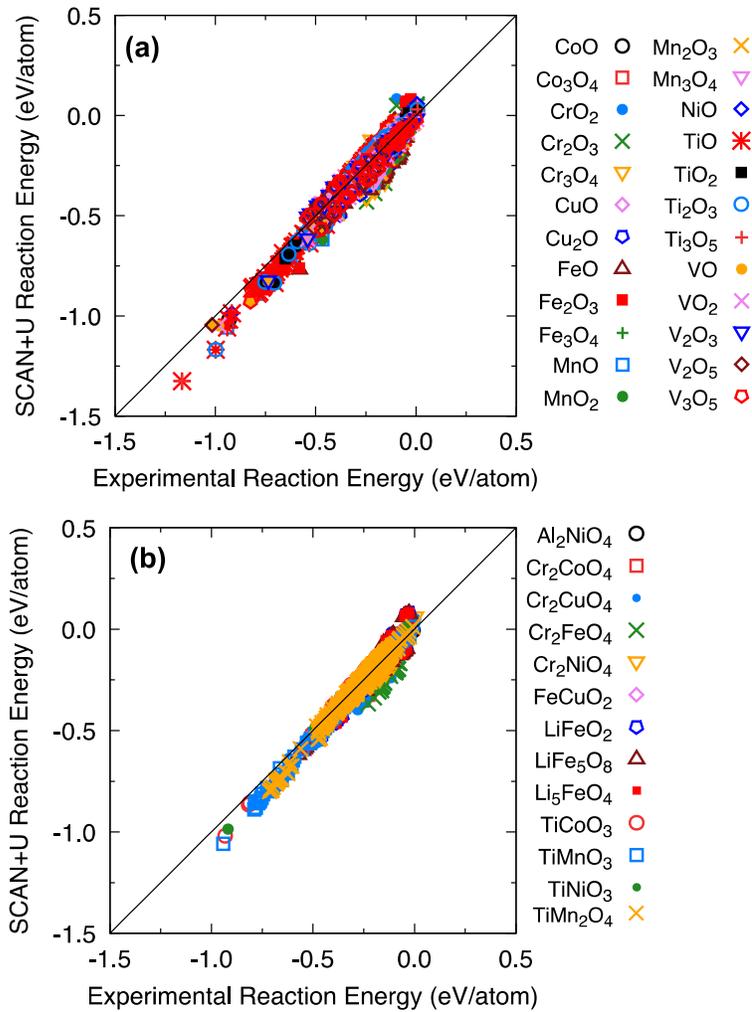

**Figure 6** Comparison of oxide reaction energies, both with and without $O_2$ as a reactant, predicted by SCAN+$U$ (L2-regularized $U$ values and Jain correction) with their reference values derived from experimental formation enthalpies for **(a)** binary oxides and **(b)** ternary oxides. Each oxide occurs in multiple reactions, so that the same symbol appears multiple times in the graphs. All data points represent true predictions from leave-one-out cross-validation.

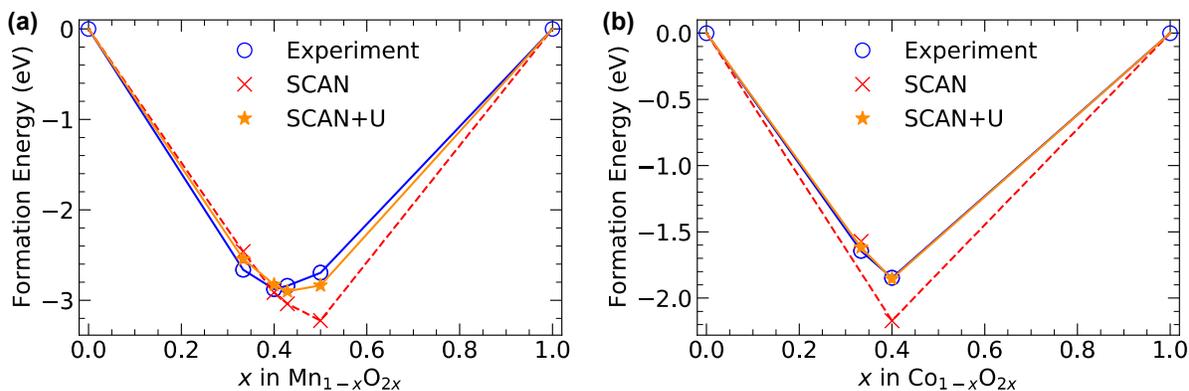

**Figure 7.** Formation energies and convex-hull constructions of the binary oxides of **(a)** Mn and **(b)** Co. Equivalent visualizations for Ti, V, Cr, and Fe are shown in supporting **Figure S5**. Formation energies computed with uncorrected SCAN+rVV10 (red crosses) and with inclusion of $L_2$-regularized optimized Hubbard-$U$ values and DFT/DFT+$U$ corrections (orange stars) are compared to experimental reference values (blue circles).

An equivalent analysis of the oxide reaction energies is shown in **Figure 6**. Note that the reaction energies vary less than the formation energies, and therefore the energy scale in **Figure 6** is smaller than that of **Figure 5**, so the scattering appears more significant. If an oxide appears in multiple reactions, its corresponding symbol appears multiple times in the graphs. Multiple compounds (reactants and products) participate in each oxide reaction, and the data points of all oxides involved in the same reaction are shown with overlapping symbols in **Figure 6**.

The largest errors are again seen for reactions involving TiO, in agreement with the error analysis of the formation energies. Note that all data points shown in **Figure 6** are true predictions obtained from LOOCV. This means, for each oxide compound in our database, a $U$ parametrization was fitted on a data set that did not include any reactions involving this specific oxide compound.

The resulting parametrization was used to predict all of the derived reaction energies. Symbols and colors in **Figure 6** indicate reaction energies that were evaluated together with the same $U$-value parametrization.

As a final test of the optimized SCAN+$U$ parameterization, we consider the phase diagram of the binary oxides that can be obtained by constructing the lower convex hull of the formation energies.[5] Only compounds that lie on the convex hull are thermodynamically stable, and the shape of the hull determines the chemical potential stability range of each phase. In **Figure 7**, the formation energy convex hulls of Mn and Co oxides as predicted by SCAN and SCAN+$U$ are compared with the experimental reference, and the corresponding constructions for Ti, V, Cr, and Fe are shown in supporting **Figure S5**. The differences between SCAN and SCAN+$U$ are most pronounced for Mn and Co with the result that SCAN+$U$ predicts the shape of the convex hulls in better agreement with experiment than uncorrected SCAN. Unlike uncorrected SCAN, SCAN+$U$ also correctly predicts both Co oxides to be stable. As seen in **Figure S5**, the differences between SCAN and SCAN+$U$ are smaller for the other transition metal species. For Ti and Fe, both SCAN and SCAN+$U$ predict the shape of the hull in good agreement with experiment but incorrectly predict some of the oxides to be slightly unstable. Both methods correctly predict all V oxides to be stable, though the +$U$ correction improves the shape of the hull. For Cr oxides, SCAN predicts two and SCAN+$U$ predicts only one of its three oxides to be stable.

4. **Discussion**

The present study considers the Hubbard-$U$ values and DFT/DFT+$U$ compatibility corrections in a unified optimization approach across the class of 3$d$ transition metal oxides. By design, the data set includes ternary oxides for which experimental formation energies are available. This results in an interconnected dataset, linking the determination of the parameters across the $3d$ series, that supports a systematic cross-validation analysis of the errors.

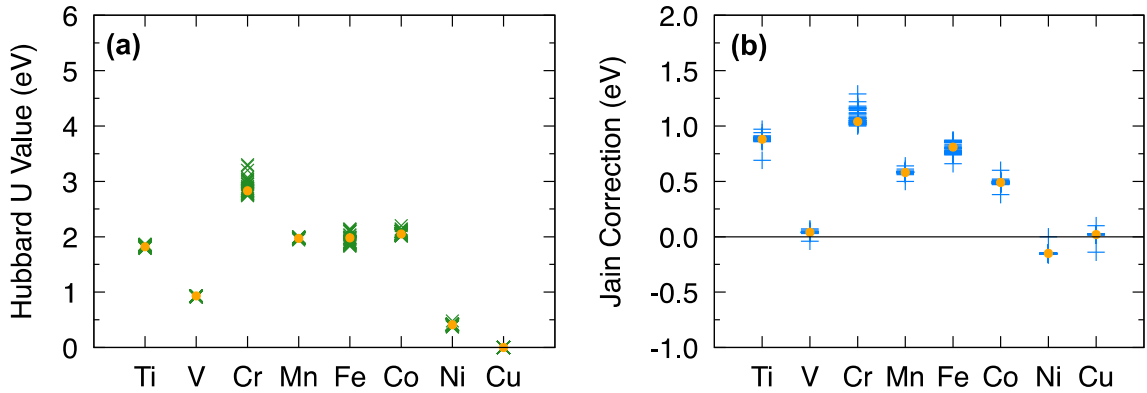

**Figure 8** Distribution of (a) $U$ values and (b) Jain correction values determined from the full set of leave-one-our cross-validation optimizations with L2 regularization ($U$+Jain+L2 parametrization). The values obtained from a single optimization on the entire data set are indicated by orange points.

We find that the SCAN functional does not exhibit any systematic error in the $O_2$ binding energy, and an empirical oxygen correction is therefore not needed. This result is in agreement with a previous study by Isaacs and Wolverton,[33] who found no systematic over- or underbinding in formation energies predicted by the SCAN functional (without Hubbard-$U$ correction).

A Hubbard-$U$ correction further reduces the error of predicted binary formation energies by more than 40% from 0.17 eV/atom to 0.10 eV/atom if a DFT/DFT+$U$ compatibility (Jain) correction is introduced as well. The error in ternary formation energies is reduced by 75% from 0.12 eV/atom to 0.03 eV/atom. Without the Jain correction, that is, only through adjustment of $U$ value, the expected errors of predicted binary and ternary formation energies are 0.12 eV/atom and 0.09 eV/atom, respectively. This is still an improvement of ~30% and 25% compared to uncorrected SCAN. Note that a Hubbard-$U$ parametrization without additional Jain correction has the advantage that it can be used with standard DFT codes without the need of additional post-processing of the predicted DFT+$U$ energies.

We note that the experimental reference data used as target for the $U$-value optimization is also subject to uncertainties. By comparing the formation enthalpies from various sources for a large number of compounds this uncertainty was previously estimated to be ~0.08 eV/atom.[65] For the present work, we found the formation enthalpies of the CRC Handbook of Chemistry and Physics[66] to be generally within ~0.02 eV/atom of the NIST-JANAF database.[67] Hence, a residual optimization error of at least this magnitude has to be expected.

**Figure 8** shows the distributions of $U$ values and Jain corrections for each TM species obtained with the LOOCV method when L2 regularization is employed. In the LOOCV procedure, the optimization is independently carried out while leaving out data deriving from one of the 37 compounds. The scatter among the 37 $U$ values and Jain corrections indicates sensitivity to data inclusion. For comparison, the values derived from a single optimization over the full dataset are also shown for each TM species. As seen in the figure, the scatter in the values is low. For all TM species, most of the optimized $U$ values lie within a range of ±0.5 eV from an average value. The values are also well centered on the single optimization results.

Note that without grouping of similar reactions and without L2 regularization, the scattering is significantly larger, leading to greater uncertainty in the $U$ values (**Figure S6**). **Table S5** lists the standard deviation of the $U$ values and Jain corrections for different optimization strategies as a measure of the scattering. The combination of reaction grouping and L2 regularization reduces the standard deviation of the $U$ values for all TM species except Fe and significantly reduces the uncertainty of the parameters for Ti and Cr. As seen in the table, the reduction is mostly due to the reaction grouping. The same trend is observed for the Jain correction values. Furthermore, although the impact of L2 regularization on the $U$ values is significant (maximum impact 2.2 eV for Cr), regularization does not noticeably affect the model accuracy, as seen in **Figure 4**. Finally, regularization results in $U$ values that are systematically smaller (**Table 1**). As a general rule, DFT+$U$ calculations are technically easier to execute, particularly to reliably converge to the correct electronic ground state, for smaller $U$ values.[26] Overall, this suggests that our use of L2 regularization has led to an improved set of Hubbard-$U$ values and DFT/DFT+$U$ compatibility corrections for application with SCAN to thermochemical properties of $3d$ TM oxide compounds.

The resulting parameters for SCAN+$U$ (with Jain correction and L2 regularization) lead to reaction energies that reproduce experiment rather accurately with an error of less than 0.08 eV RMSE across the dataset. This is comparable to the estimated errors in the underlying experimental dataset. The parameters found here also yield overall improved phase diagrams for binary oxides. However, the RMSE for a few formation and reaction energies exceeds 0.2 eV. Reactions involving titanium (II) exhibit the largest errors, and both uncorrected SCAN and SCAN+$U$ incorrectly predict TiO to be unstable (**Figure S5**). This may indicate that no single $U$ value is simultaneously appropriate for both $Ti^{2+}$ (TiO), $Ti^{3+}$ ($Ti_2O_3$), and $Ti^{4+}$ ($TiO_2$ and the ternary oxides). To a smaller degree, the same effect is seen for Fe and Mn, which also form oxides with multiple different valence states. As seen in **Figure 5**, $Fe_3O_4$ ($Fe^{2+}$ and $Fe^{3+}$) falls on the ideal diagonal while FeO ($Fe^{2+}$) and $Fe_2O_3$ ($Fe^{3+}$) are slightly over- and under-bound, respectively. Similarly, the formation enthalpy of the Mn oxide $Mn_2O_3$ ($Mn^{3+}$) is most accurately reproduced, whereas MnO ($Mn^{2+}$) and $MnO_2$ ($Mn^{4+}$) deviate in opposite directions. However, for both Fe and Mn oxides, the optimal $U$ value is an excellent compromise for all oxides and significantly improves the accuracy of reaction energies. For binary Mn oxides, SCAN+$U$ also leads to an improved phase diagram that is in excellent agreement with experiment (**Figure 7**). Without +*U* correction, SCAN incorrectly predicts two of the Fe oxides to be unstable. The +*U* correction does not correct this error but results in an overall quantitative improvement of the relative energies of the iron oxides (**Figure S5**). SCAN+*U* also improves the relative energies of the Cr oxides, however, only one out of three oxides is predicted to be stable. As noted in the introduction, the *U* value is in principle dependent on the oxidation state (*d*-band occupancy) and on the electronic screening of the local Coulomb interactions in different compounds and structures.[18] It is an approximation to assume a constant value for each TM species across all oxides. But in practical applications, the approximation of a single *U* value is essential. While a correction based on an average *U* value may not always achieve a correction of qualitative flaws in phase diagrams, such as missing phases, we find that it generally leads to a quantitative improvement of the formation energy convex hull.

Recently, Long, Gautam, and Carter reported a SCAN+*U* parameterization for the calculation of reaction energies of 3*d* TM oxides,[34,39] not including formation reactions from the elemental metals. For some of the TM species, our *U* values, which were optimized for general oxide reactions including formation reactions, differ significantly from these previous reports. Our *U* values for Fe (1.9 eV) and Ni (0.4 eV) are both significantly smaller than the values reported by Long et al. (Fe: 3.1 eV, Ni: 2.5 eV). Our optimization yields a significant *U* value for Cr (2.9 eV) whereas Long et al. found a *U* value of 0 eV (i.e., no Hubbard-*U* correction) to be optimal for oxide reactions. These differences further exemplify the importance of selecting *U* values for the intended application, an inherent limitation of DFT+*U*. Both our approach and the approach by Long et al. yield a *U* value of zero for Cu.

We note that Long et al. discuss that a $U$ value of 0 eV for Cr is counter-intuitive and appears to be an anomaly, as one might expect a non-zero optimal $U$-value for Cr that is similar in magnitude to the values for V and Mn. Long et al. attributed this behavior to error cancellation caused by potential inaccuracies in the experimental formation enthalpy of metastable CrO and the metallic character of $CrO_2$.[39] Our data set does not contain the metastable CrO. We did include $CrO_2$, as well as $Cr_2O_3$ and $Cr_3O_4$ and the stable, ternary Cr(III) spinels ($Cr_2CoO_4$, $Cr_2FeO_4$, and $Cr_2NiO_4$). Hence, the difference in $U$ value might also be due to the different choice of experimental reference.

Even though the $U$-value optimization determined that $U = 0$ eV is optimal for Cu, a small Jain correction of 0.02 eV is found to improve the overall accuracy of SCAN for Cu oxides. Hence, the correction energy is not only accounting for DFT/DFT+$U$ compatibility but also compensates other systematic errors in the formation energies. The Jain correction is only applied to oxides and only affects the formation energies. Note that alternatively an energy shift with opposite sign could be applied to the energy of the base metal to achieve an equivalent correction. Such an energy adjustment of the elements is identical to the fitted elemental-phase reference energies (FERE) by Stevanović and coworkers,[16] which is an empirical correction for improved formation enthalpy predictions.

Finally, we note that while the focus of the present work is on the prediction of formation energies and other oxide reaction energies with the SCAN functional, the optimization framework for the iterative $U$-value and Jain correction fit is general and could also be used to optimize $U$ values for other DFT functionals and for other targets, for example, to reproduce band gaps or lattice parameters. The software implementation of our approach is publicly available at https://github.com/atomisticnet/howru.

## 5. Summary and conclusions

Empirically corrected DFT calculations can yield quantitative predictions of TM oxide properties, but determining reliable and transferable parameters for DFT+$U$ methods remains a significant challenge. In this article, the parameterization of three common empirical corrections to DFT (oxygen overbinding, Hubbard-$U$, and DFT/DFT+$U$ compatibility correction) was expressed as a unified optimization problem that can be solved with the method of least squares. As an example of practical relevance, we chose to target thermochemistry of the oxides of the 3$d$ TM species (Ti, V, Cr, Mn, Fe, Co, Ni, Cu). To that purpose, we assembled a substantial database of experimental formation energies (37 compounds in total) from the literature for fitting and validation. In addition to the formation reactions, we also considered all unique reactions involving these oxides with each other and with oxygen gas, for a total of 1,710 derived reaction energies. To avoid biases in the data set, we developed a methodology for the grouping of similar reactions based on principal component analysis and $k$-means clustering. This interconnected and substantial database enabled systematic statistical evaluation of the accuracy of the derived parameters using a cross-validation methodology. In distinction over the typical procedure that focuses on one TM species at a time, this approach yields parameters for a family of elements with a procedure that is easily automated. Applied to the SCAN+$U$ density functional, we find that the error in predicted binary (ternary) oxide formation energies is reduced by ~40% (75%) if all three correction terms are included, as determined by leave-one-out cross-validation. Without a SCAN/SCAN+$U$ compatibility correction (which requires post-processing of DFT calculations with common DFT software), the improvement compared to uncorrected SCAN calculations is still ~30% (25%).

The proposed framework, which incorporates robust statistical methods, offers an approach for the simultaneous optimization of $U$ values for oxide and formation reactions. It minimizes arbitrariness in determining the empirical parameters for DFT+$U$ methods while providing an error estimate for the parameter values and the predicted reaction energies. The reported SCAN/SCAN+$U$ parameterization specifically enables accurate predictions of 3$d$ TM metal and oxide reactions. Moreover, our framework, which has been made available as free and open-source software, is not limited to the SCAN functional or to the target of reaction energetics. It can be applied to the correction of any density functional in the context of DFT+$U$ methods.

**Data availability**

The formation and reaction energy data are publicly available at

https://github.com/atomisticnet/howru.

**Code availability**

A software implementation of our approach is publicly available at

https://github.com/atomisticnet/howru.

**Acknowledgments**

This research was carried out in part at the Center for Functional Nanomaterials, which is a U.S. DOE Office of Science Facility, and used resource at the Scientific Data and Computing Center of the Computational Science Initiative, at Brookhaven National Laboratory under Contract No. DE-SC0012704. DFT calculations and machine-learning model construction also made use of the Extreme Science and Engineering Discovery Environment (XSEDE), which is supported by National Science Foundation grant number ACI-1053575 (allocation no. DMR14005). J.A.G.T. and A.U. acknowledge support by the National Science Foundation under Grant No. DMR-1940290 (Harnessing the Data Revolution, HDR).

**Author contributions**

N.A. and M.S.H. conceived the project. N.A. implemented the methodology. N.A. and J.A.G.T. performed the DFT calculations. N.A. performed the parameter optimization. N.A. and A.U. analyzed the data. N.A., A.U., and M.S.H. wrote the manuscript, which was revised by all authors.

**Ethics declarations**

*Competing interests*

The authors declare no competing interests.

# Supplementary Information:

# Data-driven Approach to Parameterize SCAN+$U$ for an Accurate Description of 3$d$ Transition Metal Oxide Thermochemistry

Nongnuch Artrith[1*†], José Antonio Garrido Torres[1], Alexander Urban[1], and Mark S. Hybertsen [2‡]

[1]Department of Chemical Engineering, Columbia University, New York, NY 10027, USA.
[2]Center for Functional Nanomaterials, Brookhaven National Laboratory, Upton, NY 11973, USA.
*Present Address: Materials Chemistry and Catalysis, Debye Institute for Nanomaterials Science, Utrecht University, 3584 CG Utrecht, The Netherlands.
[†]email: n.artrith@uu.nl [‡]email: mhyberts@bnl.gov

## 1. Supporting tables

**Table S1.** Experimental formation enthalpies ($\Delta_f H°$) at 298 K and 1 atm of the transition-metal oxides used as reference data for the $U$-value optimization. Binary oxide data taken from reference [1] unless otherwise indicated, and ternary oxide data from references [2,3].

| Binary Oxides | | Ternary Oxides | |
|---|---|---|---|
| Composition | $\Delta_f H$ (eV) | Composition | $\Delta_f H$ (eV) |
| $Co_3O_4$ | -9.234 | $Al_2NiO_4$ | -19.904 |
| $CoO$ | -2.466 | $Cr_2CoO_4$ | -14.844 |
| $Cr_2O_3$ | -11.812 | $Cr_2FeO_4$ | -14.991 |
| $Cr_3O_4$ | -15.867 | $Cr_2NiO_4$ | -14.258 |
| $CrO_2$ | -6.198 | $Cr_2CuO_4$ | -13.404 |
| $Cu_2O$ | -1.747 | $FeCuO_2$ | -5.317 |
| $CuO$ | -1.630 | $Li_5FeO_4$ | -20.400 |
| $Fe_2O_3$ | -8.542 | $LiFe_5O_8$ | -25.200 |
| $Fe_3O_4$ | -11.591 | $LiFeO_2$ | -7.7600 |
| $FeO$ | -2.819 | $TiCoO_3$ | -12.515 |
| $Mn_2O_3$ | -9.939 | $TiMnO_3$ | -14.085 |
| $Mn_3O_4$ | -14.383 | $TiNiO_3$ | -12.459 |
| $MnO$ | -3.992 | $TiMn_2O_4$ | -18.137 |
| $MnO_2$ | -5.389 | | |
| **$NiO$*** | -2.484 | | |
| $Ti_2O_3$ | -15.763 | | |
| $Ti_3O_5$ | -25.489 | | |
| $TiO$ | -5.386 | | |
| $TiO_2$ | -9.784 | | |
| $V_2O_3$ | -12.632 | | |
| $V_2O_5$ | -16.070 | | |
| $V_3O_5$ | -20.034 | | |
| $VO$ | -4.475 | | |
| **$VO_2$*** | -7.425 | | |

*From references [4,5]

**Table S2.** See separate supplementary document "Artrith-Supplementary-Reactions.pdf". Oxide reactions (in total 1,710) and their energies derived from the formation energies in **Table S1**. A data file with the reaction equations can be obtained from the GitHub repository: https://github.com/atomisticnet/howru.

**Table S3.** Experimentally measured and computed formation enthalpies of binary transition-metal and main-group oxides, as visualized for the analysis of the O–O binding energy correction in **Figure 3** of the main manuscript. Note that the enthalpies are normalized to their oxygen ($O_2$) content (in contrast to **Table S1**). Experimental data taken from reference [1] unless otherwise indicated.

| Composition | $\Delta_f H^\circ_{expt}$ (eV/$O_2$) | $\Delta_f H^\circ_{PBE}$ (eV/$O_2$) | $\Delta_f H^\circ_{SCAN}$ (eV/$O_2$) | Valence |
|---|---|---|---|---|
| $Al_2O_3$ | -11.578 | -10.073 | -11.660 | 3.0 |
| CaO | -13.160 | -11.892 | -13.266 | 2.0 |
| $Co_3O_4$ | -4.617 | -3.517 | -4.965 | 2.7 |
| CoO | -4.931 | -2.705 | -5.089 | 2.0 |
| $Cr_2O_3$ | -7.875 | -6.617 | -7.670 | 3.0 |
| $Cu_2O$ | -3.495 | -2.491 | -3.493 | 1.0 |
| CuO | -3.261 | -2.420 | -3.161 | 2.0 |
| $Fe_2O_3$ | -5.695 | -4.089 | -5.628 | 3.0 |
| $Fe_3O_4$ | -5.796 | -4.040 | -5.711 | 2.7 |
| FeO | -5.638 | -4.049 | -5.576 | 2.0 |
| $Li_2O$ | -12.393 | -11.223 | -12.489 | 1.0 |
| MgO | -12.470 | -10.851 | -12.458 | 2.0 |
| $Mn_2O_3$ | -6.626 | -5.517 | -6.592 | 3.0 |
| $Mn_3O_4$ | -7.192 | -5.485 | -7.172 | 2.7 |
| MnO | -7.984 | -4.942 | -8.063 | 2.0 |
| $MnO_2$ | -5.389 | -5.025 | -5.379 | 4.0 |
| $Na_2O$ | -8.586 | -7.318 | -8.737 | 1.0 |
| **NiO**\* | -4.968 | -2.405 | -4.882 | 2.0 |
| $Sc_2O_3$ | -13.189 | -11.891 | -13.212 | 3.0 |
| $SiO_2$ | -9.438 | -8.462 | -9.400 | 4.0 |
| $SnO_2$ | -5.986 | -4.947 | -5.959 | 4.0 |
| TiO2 | -9.784 | -9.159 | -9.753 | 4.0 |
| $V_2O_3$ | -8.421 | -7.637 | -8.312 | 3.0 |
| $V_2O_5$ | -6.428 | -6.459 | -6.001 | 5.0 |
| VO | -8.950 | -6.595 | -8.799 | 2.0 |
| **$VO_2$**\* | -7.425 | -7.129 | -7.317 | 4.0 |
| ZnO | -7.265 | -5.768 | -7.220 | 2.0 |

\*From references [4,5]

**Table S4.** Root mean squared errors in electronvolts per atom for oxide formation and oxide/$O_2$ reaction energies predicted by SCAN+$U$ calculations using the three different sets of $U$ values and Jain corrections of **Table 1** in the main manuscript. All SCAN+$U$ error estimates were obtained from leave-one-out cross validation.

| Unit: eV/atom | No $U$ | $U$ only | $U$ only (L2) | $U$+Jain | $U$+Jain (L2) |
|---|---|---|---|---|---|
| Binary oxide formation | 0.17 | 0.13 | 0.12 | 0.11 | 0.10 |
| Ternary oxide formation | 0.12 | 0.09 | 0.09 | 0.03 | 0.03 |
| Binary oxide reactions | 0.05 | 0.05 | 0.05 | 0.05 | 0.05 |
| Ternary oxide reactions | 0.06 | 0.03 | 0.03 | 0.02 | 0.02 |
| Binary $O_2$ reactions | 0.14 | 0.08 | 0.08 | 0.07 | 0.07 |
| Ternary $O_2$ reactions | 0.05 | 0.04 | 0.04 | 0.03 | 0.03 |

**Table S5.** Standard deviation of the Hubbard-$U$ values and the Jain corrections obtained from leave-one-out cross-validation for different optimization strategies: Direct least-squares optimization (Direct), L2-regularized optimization (L2 regularized), optimization with grouping of similar reactions using principal component analysis and *k*-means clustering (Reaction grouping), and optimization with both L2 regularization and reaction grouping (Grouping & L2). As seen in the data, the reaction grouping significantly reduces the magnitude of the standard deviation for those transition-metal species with large initial error bar (Ti, Cr). For all species except Fe, the uncertainty in the $U$ values decreases when reaction grouping and L2 regularization are employed. The trends in the standard deviations of the Jain correction values follow those of the $U$ values, since the Jain corrections are determined for each set of $U$ values.

| | Ti | V | Cr | Mn | Fe | Co | Ni | Cu |
|---|---|---|---|---|---|---|---|---|
| ***U* Value Standard Deviation (eV)** | | | | | | | | |
| Direct | 0.46 | 0.10 | 0.30 | 0.10 | 0.09 | 0.11 | 0.03 | 0.00 |
| L2 regularization | 0.26 | 0.09 | 0.27 | 0.11 | 0.10 | 0.18 | 0.10 | 0.00 |
| Reaction grouping | 0.03 | 0.03 | 0.14 | 0.01 | 0.08 | 0.03 | 0.07 | 0.00 |
| Grouping & L2 | 0.02 | 0.05 | 0.14 | 0.01 | 0.10 | 0.04 | 0.01 | 0.00 |
| **Jain Correction Standard Deviations (eV)** | | | | | | | | |
| Direct | 0.27 | 0.06 | 0.16 | 0.04 | 0.04 | 0.04 | 0.01 | 0.00 |
| L2 regularization | 0.15 | 0.05 | 0.14 | 0.04 | 0.04 | 0.06 | 0.03 | 0.00 |
| Reaction grouping | 0.04 | 0.04 | 0.07 | 0.01 | 0.03 | 0.01 | 0.02 | 0.00 |
| Grouping & L2 | 0.04 | 0.01 | 0.07 | 0.02 | 0.05 | 0.03 | 0.02 | 0.00 |

## 2. Supporting figures

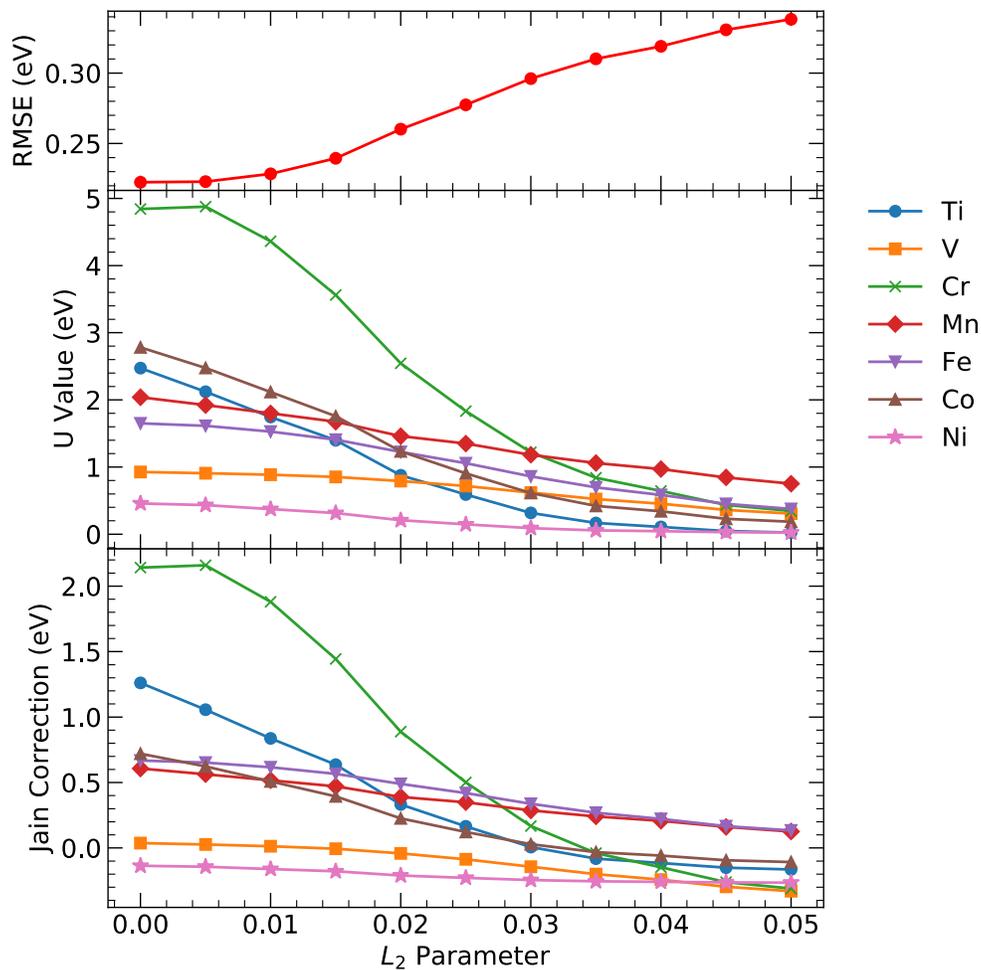

**Figure S1.** Impact of L2 regularization with different regularization parameters, λ of equation (13) in the main manuscript, on the root mean square error (RMSE) of the reaction energies (top), the $U$ values (middle), and the DFT/DFT+$U$ compatibility correction after Jain (bottom).

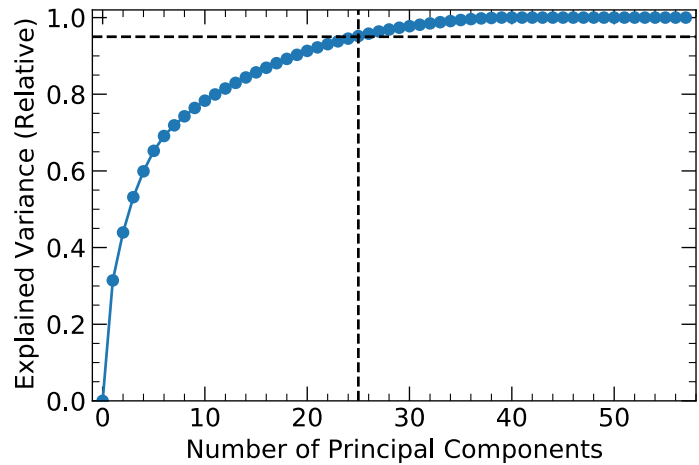

**Figure S2.** Explained variance as a function of the number of principal components (PCs) in a PC analysis of the 1,710 reactions in the reference data. The dashed line indicates 25 PCs.

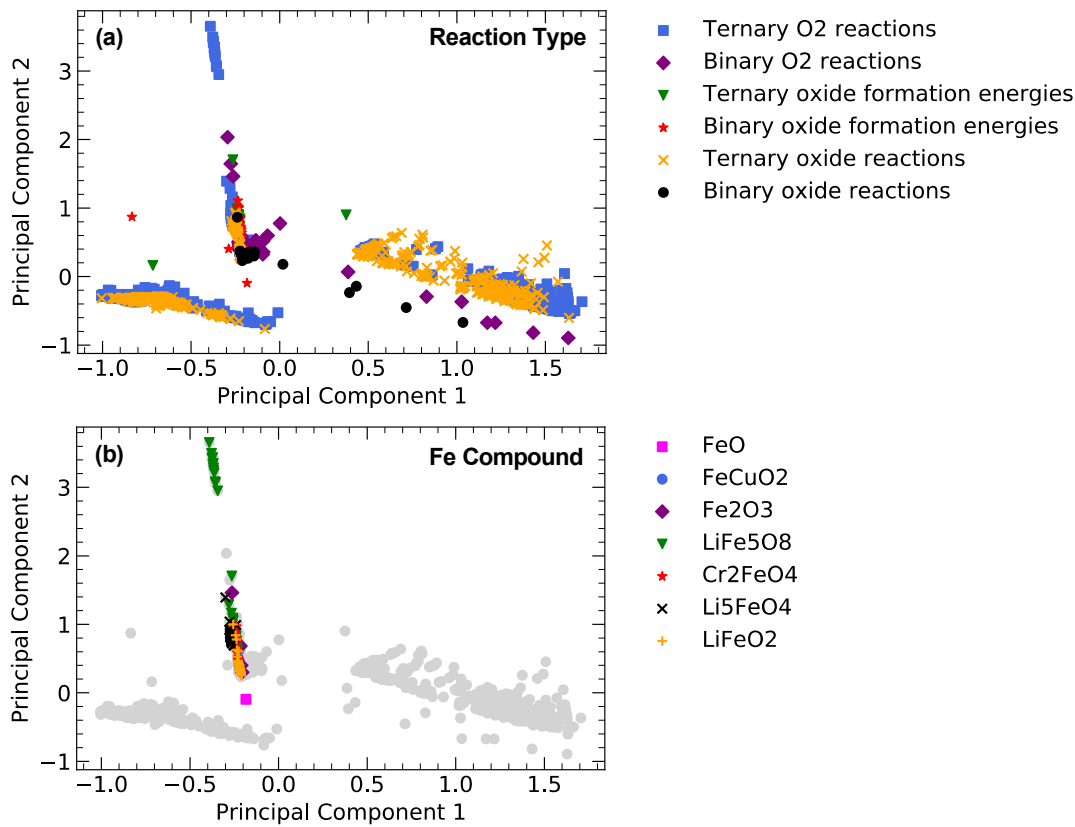

**Figure S3.** Visualization of the reaction data set in the coordinates of the first two principal components (PCs) from a PC analysis. Reactions colored by (a) reactions type and (b) reaction product (for the example of Fe compounds).

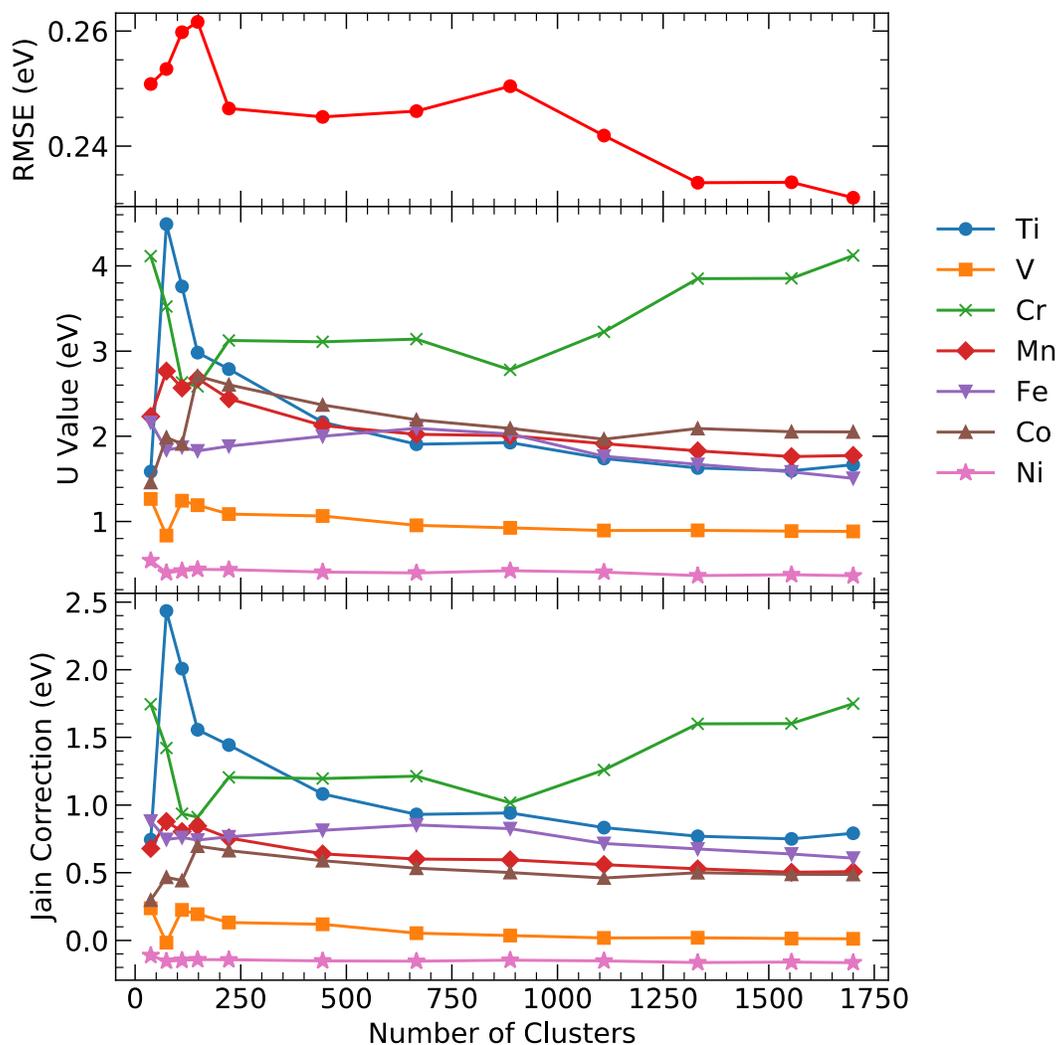

**Figure S4.** Dependence of the root mean squared error (RMSE) of the predicted reaction energies, the $U$ values, and the DFT/DFT+$U$ compatibility correction after Jain on the number of clusters in $k$-means clustering.

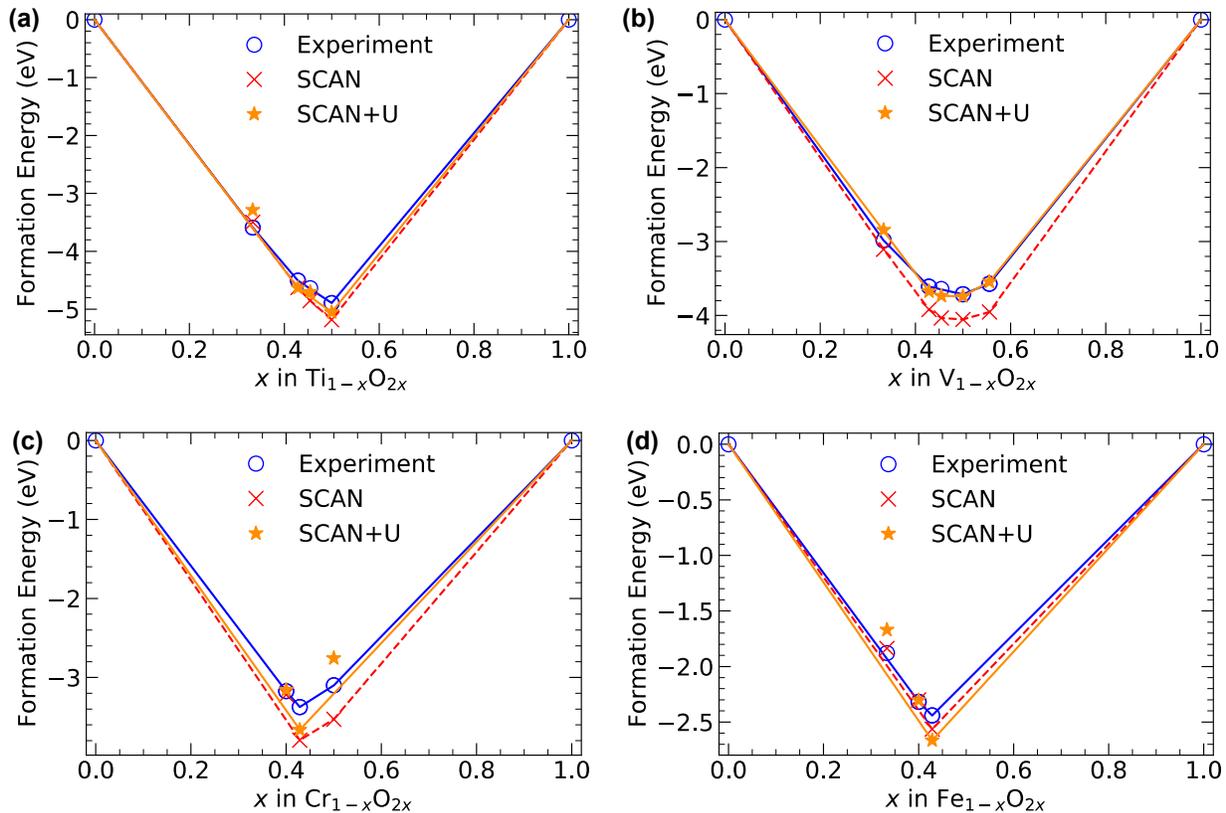

**Figure S5.** Formation-energies and convex hull constructions for the binary oxides of (a) Ti, (b) V, (c) Cr, and (d) Fe. The corresponding graphs for Mn and Co are shown in **Figure 7** of the main manuscript. Formation energies obtained from uncorrected SCAN+rVV10 calculations are indicated by red crosses, and orange stars are the formation energies calculated with $L_2$-regularized optimized Hubbard-$U$ values and DFT/DFT+$U$ corrections. The corresponding experimental reference values are shown as blue circles.

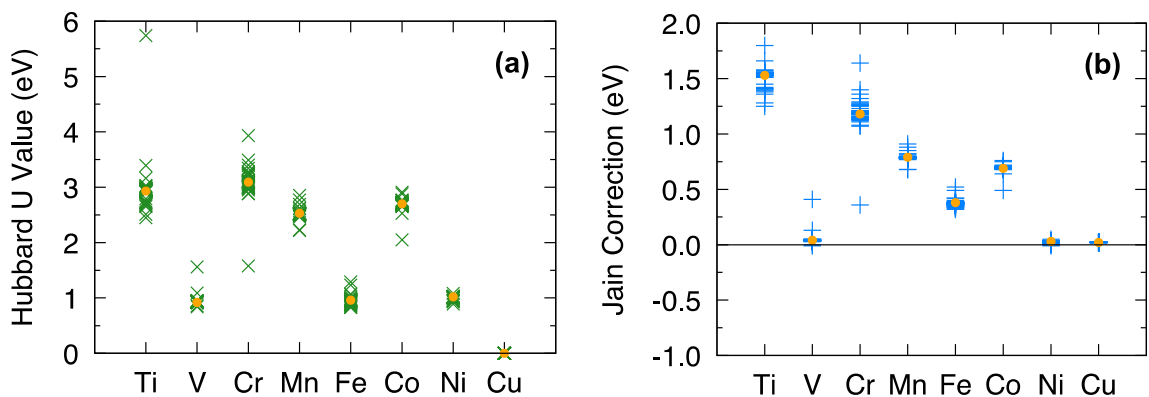

**Figure S6.** *U*+Jain **parametrization without reaction grouping and L2 regularization:** Distribution of **(a)** *U* values and **(b)** DFT/DFT+*U* compatibility (Jain) correction values during leave-one-our cross-validation without grouping of similar reactions and without L2 regularization. The values obtained from optimization on the entire data set are indicated by orange points. Reaction grouping reduces the amount of scattering, as seen in **Figure 7** of the main manuscript.